\documentclass[pdflatex,sn-mathphys-num]{sn-jnl}
\usepackage[utf8]{inputenc}

\usepackage{graphicx}%
\usepackage{multirow}%
\usepackage{amsmath,amssymb,amsfonts}%
\usepackage{amsthm}%
\usepackage{mathrsfs}%
\usepackage[title]{appendix}%
\usepackage{xcolor}%
\usepackage{textcomp}%
\usepackage{manyfoot}%
\usepackage{booktabs}%
\usepackage{algorithm}%
\usepackage{algorithmicx}%
\usepackage{algpseudocode}%
\usepackage{listings}%
\usepackage{amssymb}
\usepackage{amsmath}
\usepackage{url}    
\usepackage{color,soul}
\usepackage{xr}
\usepackage{cleveref}
\usepackage{multirow}
\theoremstyle{thmstyleone}%
\theoremstyle{thmstyletwo}%

\theoremstyle{thmstylethree}%
\begin{document}
\title[Article Title]{Integrated Open-Source Framework for Simulation of Transcatheter Pulmonary Valves in Native Right Ventricular Outflow Tracts}


\author[1]{Christopher N. Zelonis}
\author*[1]{\fnm{Jalaj Maheshwari}}\email{maheshwarj@chop.edu}
\author[2,3,6]{Wensi Wu}
\author[4]{Steve A. Maas}
\author[1]{Seda Aslan}
\author[5]{Kyle Sunderland}
\author[1]{Stephen Ching}
\author[1]{Ashley Koluda}
\author[1]{Yuval Barak-Corren}
\author[1]{Nicolas Mangine}
\author[1]{Patricia M. Sabin}
\author[5]{Andras Lasso}
\author[1,6]{Devin W. Laurence}
\author[1]{Christian Herz}
\author[6]{Matthew J. Gillespie}
\author[4]{Jeffrey A. Weiss}
\author[1,6]{Matthew A. Jolley}

\affil[1]{\orgdiv{Department of Anesthesiology and Critical Care Medicine}, \orgname{Children's Hospital of Philadelphia}, \city{Philadelphia}, \postcode{19104}, \state{PA}, \country{USA}}

\affil[2]{\orgdiv{Cardiovascular Institute}, \orgname{Children's Hospital of Philadelphia}, \city{Philadelphia}, \postcode{19104}, \state{PA}, \country{USA}}

\affil[3]{\orgdiv{Department of Mechanical Engineering and Applied Mechanics}, \orgname{University of Pennsylvania}, \city{Philadelphia}, \postcode{19104}, \state{PA}, \country{USA}}

\affil[4]{\orgdiv{Scientific Computing and Imaging Institute}, \orgname{University of Utah}, \city{Salt Lake City}, \postcode{84112}, \state{UT}, \country{USA}}

\affil[5]{\orgdiv{Laboratory for Percutaneous Surgery}, \orgname{Queen’s University}, \city{Kingston}, \postcode{K7L 2N8}, \state{ON}, \country{Canada}}

\affil[6]{\orgdiv{Division of Pediatric Cardiology}, \orgname{Children's Hospital of Philadelphia}, \city{Philadelphia}, \postcode{19104}, \state{PA}, \country{USA}}


\abstract{Pulmonary insufficiency is a consequence of transannular patch repair in Tetralogy of Fallot (ToF), leading to late morbidity and mortality.  Transcatheter native outflow tract pulmonary valve replacement (TPVR) has become common, but assessment of patient candidacy and selection of the optimal device remains challenging. We demonstrate an integrated open-source workflow for simulation of TPVR in image-derived models to inform device selection. Machine learning-based segmentation of Computer Tomography (CT) scans was implemented to define the right ventricular outflow tract (RVOT).  A custom workflow for device positioning and pre-compression was implemented in SlicerHeart. Resulting geometries were exported to FEBio for simulation.  Visualization of results and quantification were performed using custom metrics implemented in SlicerHeart and FEBio. RVOT model creation and device placement could be completed in under 1 minute. Virtual device placement using FE simulations visually mimicked actual device placement and allowed quantification of vessel strain, stress, and contact area. Regions of higher strain and stress were observed at the proximal and distal end locations of the TPVs where the devices impinge on the RVOT wall. No other consistent trends were observed across simulations. The observed variability in mechanical metrics across RVOTS, stents, and locations in the RVOT highlights that no single device performs optimally in all anatomies, thereby reinforcing the need for simulation-based patient-specific assessment. This study demonstrates the feasibility of a novel open-source workflow for the rapid simulation of TPVR, which, with further refinement, may inform assessment of patient candidacy and optimal device selection.}




\maketitle

\section{Abbreviations}
    
\textbf{2D} Two Dimensional \\
\textbf{3D} Three Dimensional \\
\textbf{CHD} Congenital Heart Disease \\
\textbf{CTA} Computed Tomography Angiography \\
\textbf{DICOM} Digital Imaging and Communications in Medicine \\
\textbf{FE} Finite Element \\
\textbf{HGO} Holzapfel Gasser Ogden \\
\textbf{RVOT} Right Ventricular Outflow Tract \\
\textbf{RV-PA} Right Ventricle to Pulmonary Artery \\
\textbf{STL} Stereolithography \\
\textbf{ToF} Tetralogy of Fallot \\
\textbf{TPV} Transcatheter Pulmonary Valve \\
\textbf{TPVR} Transcatheter Pulmonary Valve Replacement \\

\section{Introduction}\label{sec1}

Tetralogy of Fallot (ToF) is a form of congenital heart disease (CHD) characterized by pulmonary valve stenosis, right ventricular hypertrophy, a ventricular septal defect, and an overriding aorta. ToF is the most common form of cyanotic CHD, with a prevalence of about 5 out of every 10,000 live births \cite{stallings_national_2024}. In most cases of ToF, a surgical patch is placed to enlarge the pulmonary valve annulus (transannular patch). This patch relieves the obstruction to blood flow in infancy but results in loss of competence of the pulmonary valve.  While initially tolerated, over time, severe regurgitation of the pulmonary valve results in right ventricular dilation, ventricular failure, and life-threatening arrhythmias \cite{geva_preoperative_2018, bokma_improved_2023, bokma_propensity_2018, valente_contemporary_2014}. Surgical implantation of a bioprosthetic valve has been used to remedy pulmonary insufficiency and related complications, but has the associated morbidity of open heart surgery on cardiopulmonary bypass \cite{law_transcatheter_2021, gillespie_1-year_2023, crago_pediatric_2023, gartenberg_transcatheter_2023, jolley_toward_2019}.

Over the last two decades, transcatheter pulmonary valve replacement (TPVR) has evolved to avoid the risks of open-heart surgery \cite{gartenberg_transcatheter_2023, jolley_toward_2019, kenny_current_2017}. The first transcatheter pulmonary valve (TPV) devices available, the Melody valve (Medtronic, Minneapolis, MN) and Sapien valve (Edwards Lifesciences, Irvine, CA), were balloon-expandable bioprosthetic valves approved only for implantation in the small percentage of patients who received surgically placed cylindrical right ventricle-to-pulmonary artery (RV-PA) conduits as part of their initial repair \cite{gartenberg_transcatheter_2023, jolley_toward_2019, patel_transcatheter_2022}. However, the vast majority of patients with repaired ToF have surgically patched native outflow tracts, which demonstrate significant anatomic variability \cite{schievano_variations_2007, odemis_3d_2025}. As such, nearly 80\% of patients with ToF are not candidates for traditional balloon expandable TPV therapies \cite{gillespie_1-year_2023, gartenberg_transcatheter_2023, jolley_toward_2019, kenny_current_2017, patel_transcatheter_2022, schievano_variations_2007, capelli_patient-specific_2010}.

In order to meet this need, self-expanding TPV systems that can conform to a wide range of heterogeneous native RVOTs have been developed. In contrast to balloon-expandable TPVs, self-expanding TPVs consist of a spring-like Nitinol frame enclosing a tissue valve \cite{benson_three-year_2020, zahn_first_2018, jin_five-year_2024}, which can expand to conform to the vessel and facilitate device retention. However, the precise fit of the device in a given individual is critical; if the device is dislodged, it can catastrophically occlude the pulmonary outflow. Further, if the proximal and distal regions of the device do not effectively seal the pulmonary artery, a paravalvular leak or device thrombosis may result \cite{benson_three-year_2020, schoonbeek_implantation_2016, gillespie_patient_2017}.  

To date, the anatomical suitability of  TPVR for a given patient is performed by the vendor using a screening CT scan of the RVOT \cite{mcelhinney_transcatheter_2024}.  For example, the screening process of the  Harmony valve (Medtronic) was developed as part of the early feasibility study and subsequently refined during subsequent trials based on an engineering analysis that estimated degree of device oversizing (or interference) needed to secure the device in a given RVOT \cite{benson_three-year_2020, gillespie_patient_2017, bergersen_harmony_2017}.  This analysis is now performed by measuring the circumference of the RVOT at different levels and overlaying the device shape on the perimeter-derived radius of the vessel, resulting in a “perimeter plot” \cite{mcelhinney_transcatheter_2024}.  However, this simplification of the anatomic geometry does not capture the true physical interaction of the device with the complex shape of the RVOT, and fails to account for vessel expansion in the setting of device placement \cite{jolley_toward_2019}. In addition, many patients are not determined to be candidates for TPVR by standard criteria, but the sensitivity and specificity of these relatively simple screening processes are unknown \cite{mcelhinney_transcatheter_2024}. As such, patients may be excluded unnecessarily due to conservative guidelines. Adding to the challenge, there are 3 different devices from two different manufacturers currently approved in the United States and over 15 different TPVs across a range of designs and sizes in use or in development internationally \cite{benson_three-year_2020, zahn_first_2018, jin_five-year_2024}.  As the number of available devices for TPVR from a range of vendors increases, there is a growing need for an accurate method for assessing patient candidacy and matching the optimal device to an individual patient. In this setting, a more physically representative, patient-specific geometrical compatibility method to select the optimal device for an individual patient within a highly heterogeneous population is of increasing relevance. 

Several groups have explored the modeling of TPVR over the last two decades, which laid the groundwork for the current generation of patient-specific simulations. Capelli et al.\cite{capelli_patient-specific_2010, capelli_patient-specific_2012} pioneered early simulations of self-expanding devices in image-derived RVOT models, laying the foundation for later studies. Their early implementations assumed a rigid, non-deformable RVOT, now recognized as a significant simplification \cite{jolley_toward_2019}. Bosi et al. \cite{bosi_patientspecific_2015} advanced the field by introducing deformable vessel behavior and simulating both balloon-expandable and self-expanding valves using Abaqus, though with simplified isotropic linear elastic models for the vessel wall. Further advancements have aimed to improve the physiological accuracy of vessel mechanics. Donahue et al. \cite{donahue_finite_2022}, for example, applied the Holzapfel-Gasser-Ogden (HGO) constitutive model to account for anisotropy and fiber orientation in RVOT tissue. However, their work focused on simulating balloon expansion through wall pressurization to assess coronary compression risk, rather than modeling the deployment of a self-expanding TPV device. Notably, these foundational simulations predated the commercial availability of dedicated self-expanding TPV devices for pediatric patients.

Patients with repaired congenital heart disease represent a small and anatomically heterogeneous population for whom commercially available modeling and screening tools are often poorly suited, as these technologies are typically developed for larger markets where commercial incentives are greatest. Yet these are precisely the patients for whom device fit is most difficult to predict, for whom no single TPV device is universally optimal, and for whom anatomic suitability is currently assessed through proprietary, single-vendor workflows that cannot be independently examined, reproduced, or compared across devices. Clinicians are increasingly implanting these devices even when standard fit criteria are not met \cite{medtronic_tpv_2021}, underscoring the need for more physically representative, patient-specific approaches to device selection and procedural planning. 

To address this gap, we demonstrate the creation and initial application of a complete open-source workflow for finite element (FE) simulation of self-expanding TPV deployment in image-derived, patient-specific RVOT geometries, with the long-term goal of informing both patient candidacy and optimal device selection. We implemented this capability within the SlicerHeart extension for 3D Slicer and FEBio \cite{fedorov_3d_2012, lasso_slicerheart_2022, maas_febio_2012}. Specific contributions include rapid machine learning-based segmentation of the RVOT and coronary arteries, a novel interface for TPV selection and positioning within the RVOT, and direct export from SlicerHeart to FEBio for simulation. We further demonstrate virtual implantation of multiple TPV devices across a range of RVOT morphologies and quantify device shape, compression, and vessel contact to inform device selection and placement in individual patients. Beyond its technical capabilities, this open framework promotes access by allowing any investigator or center to inspect and apply the methodology without proprietary software; advances equity by enabling direct comparison across devices and vendors; and fosters innovation by providing a shared, extensible foundation that can be scrutinized and improved by the broader community rather than recreated within isolated, closed pipelines. With further development and validation, this workflow could support personalized TPVR planning and be generalized to other self-expanding medical devices.

\section{Methods}\label{sec2}

\begin{figure}
    \centering
    \includegraphics[width=1.0\linewidth]{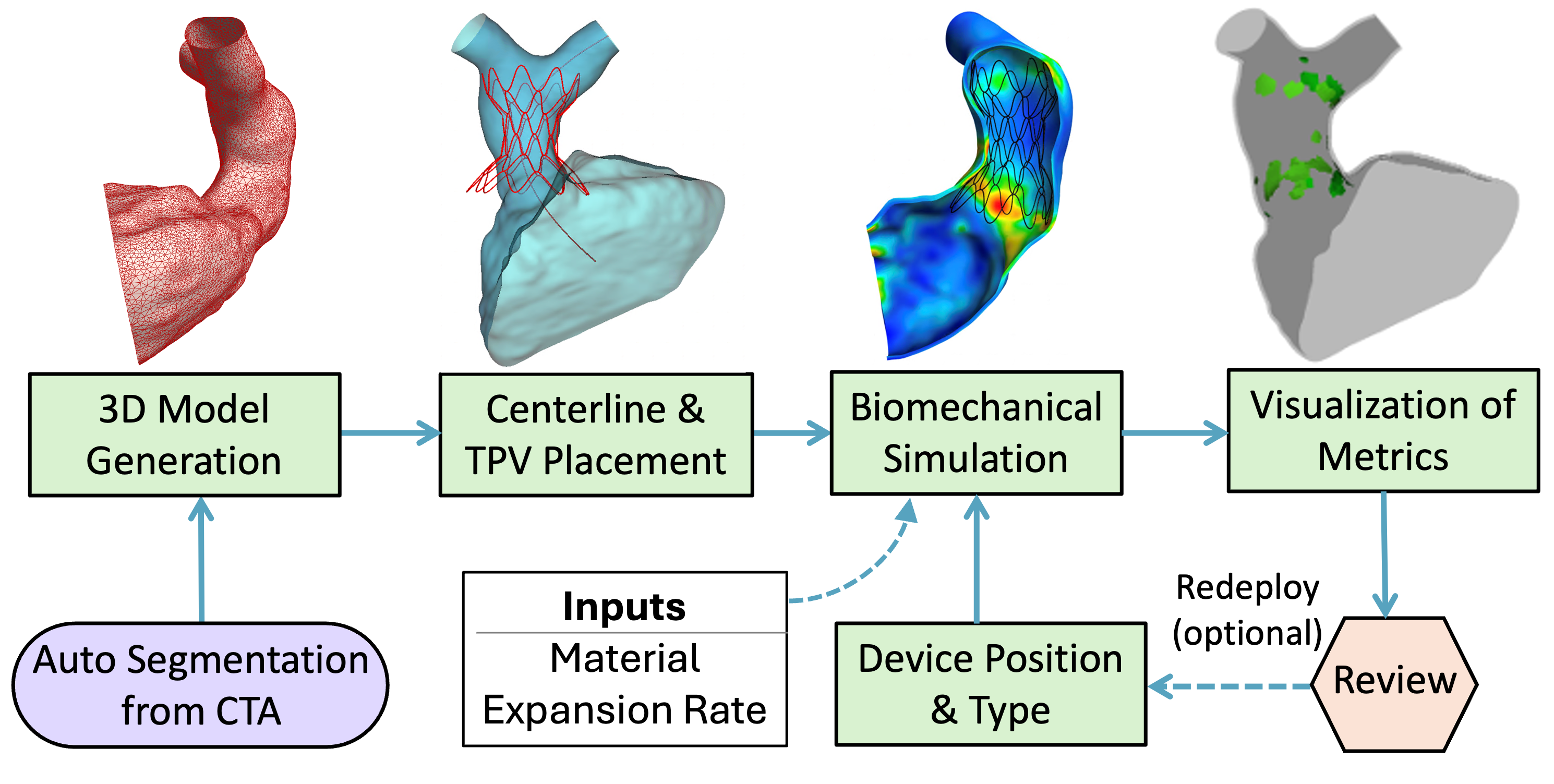}
    \caption{Overview of Modeling Pipeline}
    \label{fig:workflow}
\end{figure}

The modules and other code used for image segmentation are included in the SlicerHeart extension for 3D Slicer and are also available at \url{www.github.com/SlicerHeart} \cite{fedorov_3d_2012, lasso_slicerheart_2022}. The open-source finite element solver, FEBio version 4.7.0, was used for all FE simulations of TPV deployment \cite{maas_febio_2012}. The modeling workflow utilized is displayed in Figure \ref{fig:workflow}. A more detailed workflow figure outlining the software used, inputs, and outputs to each step is included in the supplementary material as Figure A1.

\subsection{Subjects}
An existing institutional database was used to retrospectively identify patients with a diagnosis of ToF who underwent TPVR and in whom computed tomography angiography (CTA) of the RVOT prior to TPVR had been previously acquired. Three patients with ToF (21.19, 16.40, and 13.72 years old) were chosen to demonstrate device deployment in RVOT anatomies of varying sizes and shapes. TPVR is typically applied in children from 8 years to adults, with a median age between 19 and 26 years. Other studies have reported TPVR generally performed for patients in a wide age range, which includes children, adolescents, young adults, and older adults \cite{nordmeyer_acute_2019, armstrong_association_2019, houeijeh_long-term_2023}. The Institutional Review Board at the Children’s Hospital of Philadelphia approved this study.

\subsection{Image Acquisition}
A retrospective ECG-gated dual-source scanner (Siemens Healthineers, Forchheim, Germany) was used for image acquisition, with 2 x 128 x 0.6 mm slice collimation. 350 mg/mL Omnipaque (iohexol, GE Healthcare, Chicago, IL), a low-osmolar iodinated contrast, was injected via peripheral intravenous access at a dose of 2 mL/kg, up to 100 mL. The contrast was followed by an injection of normal saline using a dual-head power injector (Medrad Inc, Warrendale, PA). CTA images were available only for the diastolic phase for all three patients explored in this study. CTA images were imported into 3D Slicer (\url{www.slicer.org}) in Digital Imaging and Communications in Medicine (DICOM) format. CT images of three unconstrained TPV devices were acquired on the CT scanner to assist in the creation of realistic device models. In addition, measurements and dimensions were confirmed with Neiko 01407A digital calipers sensitive to 0.01 mm (Zhejiang Kangle Group, Wenzhou, China). More details of the TPV device designs are provided in Section 3.4.1.

\subsection{RVOT Segmentation}
Conversion of RVOT CT scans to segmented models has been extensively described in our previous work \cite{barak-corren_image-derived_2025}. A brief description of the process is as follows. Initial segmentations were generated using threshold paint with 3D sphere brushes and manual revision using the Segment Editor module in 3D Slicer as previously described \cite{barak-corren_image-derived_2025}. The sphere brushes were given a diameter of 2\% of the screen size for changes that span multiple slices and 1\% for finer edits on a single slice view. A deep learning-based model was trained to automatically segment the RVOT and coronary arteries, as previously described \cite{barak-corren_image-derived_2025}. MONAI’s DeepEdit was modified for this population \cite{nguyen_deepedit_2022} as we recently described \cite{barak-corren_image-derived_2025}. During model training, the CT images went through a number of transforms for normalization and data augmentation to introduce more variation. All images were resampled to a resolution of 1mm x 1mm x 1mm. During preliminary observations, the intensity values of RVOT and coronary arteries were located in the range 125-550 HU. The image intensities were thresholded to 125 $\leq$ I $\leq$ 550 HU and scaled to 0.0 $\leq$ I $\leq$ 1.0, with I representing the image intensity. Data augmentation included random flips, intensity shifts, and 90-degree rotations. Finally, datasets were centrally cropped/padded to have a spatial size of 192 x 192 x 192.

\begin{figure}
    \centering
    \includegraphics[width=1.0\linewidth]{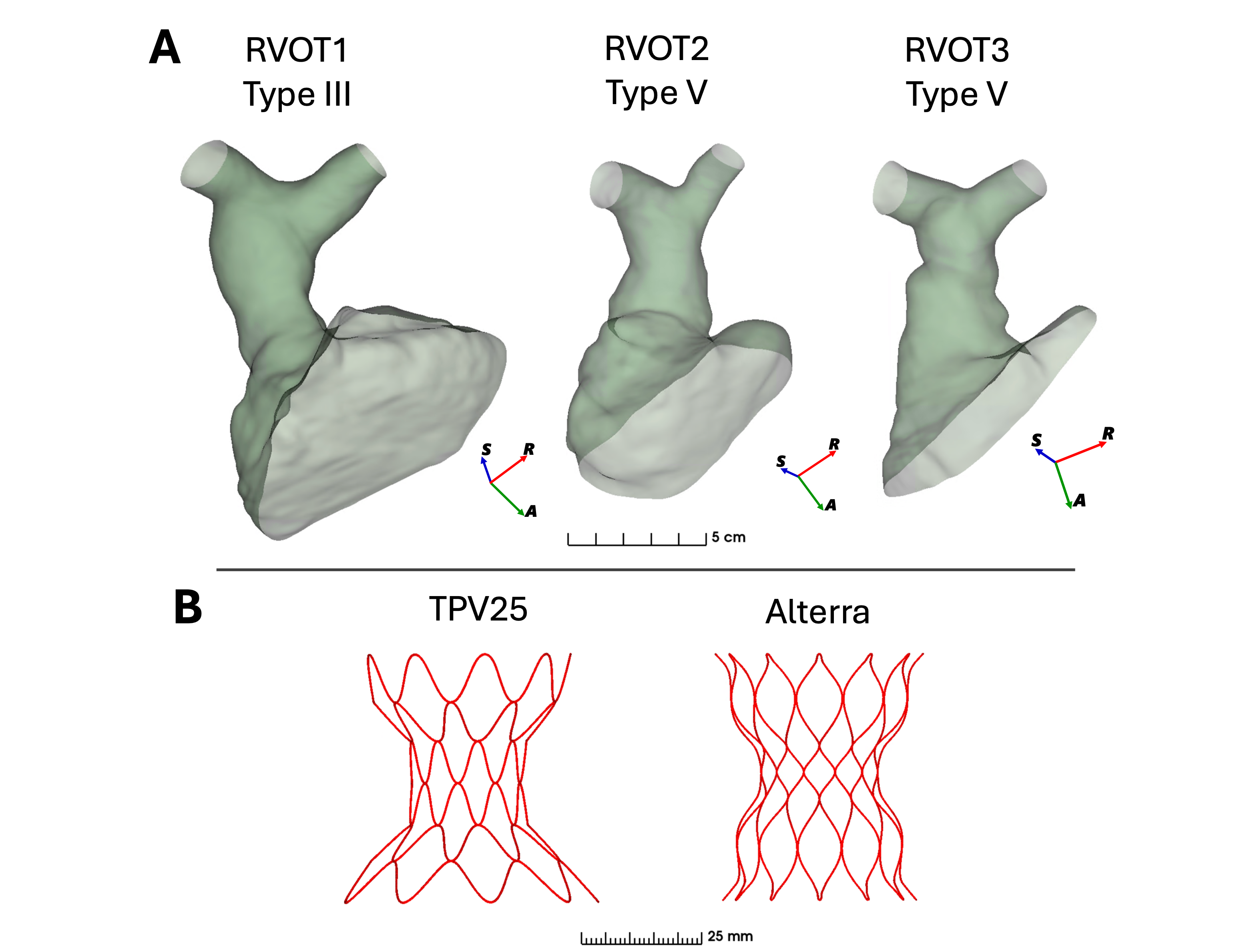}
    \caption{A) Shell models of our 3 patient-specific RVOT vessels, to scale. The example vessels were chosen to demonstrate workflow applicability in a range of RVOT sizes and shapes. RVOT morphopolgy types are determined as per \cite{schievano_variations_2007}. R, S, and A axes denote the right, superior, and anterior directions. B) TPV device models available in the TPVR Simulation module within SlicerHeart.}
    \label{fig:RVOTs_and_TPVs}
\end{figure}

Training was performed on 80 datasets for 50 epochs. As training advanced, the deep-learning model was utilized for the creation of further training samples with manual inspection and, if necessary, modification by a physician with prior segmentation experience. 100 models were created in total. Segmentations of the RVOT examples herein were made with deep learning and the MONAILabel module in 3D Slicer \cite{diaz-pinto_monai_2024}. The segmentation models produced by this module may need manual intervention before they can be used for the subsequent steps in the workflow. Unintended areas may be captured due to non-homogeneous blood pools sometimes observed in CT scans. Areas in the imaging that are very dark or light may not be captured well and may appear fuzzy in the segmentation. This fuzziness needs to be smoothed manually before proceeding to the next step. The RVOT segmentations used in this study were subsequently cleaned and smoothed in the Segment Editor module. RVOT segmentations were converted to models and the ends were uncapped using the Dynamic Modeler tool to form hollow models. The automatic segmentation process is shown in Video-1. Figure \ref{fig:RVOTs_and_TPVs} shows the segmented RVOTs used for this study. The respective RVOT morphology types based on perimeter-derived diameters at different lengths along the RVOT \cite{schievano_variations_2007} are also highlighted in \ref{fig:RVOTs_and_TPVs}.

\subsection{TPVR Simulation Module Features}

\subsubsection{Device Design}
A custom TPVR Simulation module was developed in SlicerHeart to allow placement of a few widely available, commercial TPV devices into the RVOT \cite{lasso_slicerheart_2022}. The module currently supports positioning of the Harmony TPV-22 (Medtronic, Minneapolis, MN), Harmony TPV-25 (Medtronic, Minneapolis, MN), and the Alterra adaptive prestent (Edwards Lifesciences, Irvine, CA). Models of these devices were created in the Computer Aided Design software, Fusion 360 (Autodesk, San Francisco, CA), from CT imaging and measurements specified in the device documentation \cite{medtronic_tpv_2021, edwards_lifesciences_edwards_2021}. Only the TPV-25 and Alterra devices were used in this study since they are the most used clinically. Additionally, while the functionality to use the TPV-22 device is available, the discontinuity in the metal frame between the upper and lower device introduces additional complexity relative to the singular frame of the TPV-25 and Alterra devices. The TPV devices are made of Nitinol with shape-memory capabilities. The TPV-25 wireframe devics was modeled with a wire diameter of 0.375 mm. The true Alterra device consists of a laser-cut frame; each wire-like piece has a rectangular cross-section with rounded edges and approximate dimensions of 0.25 mm by 0.32 mm. We modeled the Alterra with a round wire of diameter 0.32 mm for simplicity. We based this value on the diameter of a circle that yields the same cross-sectional area as the actual rectangular cross-section of the Alterra. The TPVs were remeshed as tetrahedral elements using TetWild \cite{hu_tetrahedral_2018} and GMSH \cite{geuzaine_gmsh_2009}, then integrated directly into the TPVR Simulation module as .vtk files for easy user selection. Figure \ref{fig:RVOTs_and_TPVs} also shows the TPV devices used in this study.

\subsubsection{Device Positioning in Right Ventricular Outflow Tract}
\begin{figure}
    \centering
    \includegraphics[width=1.0\linewidth]{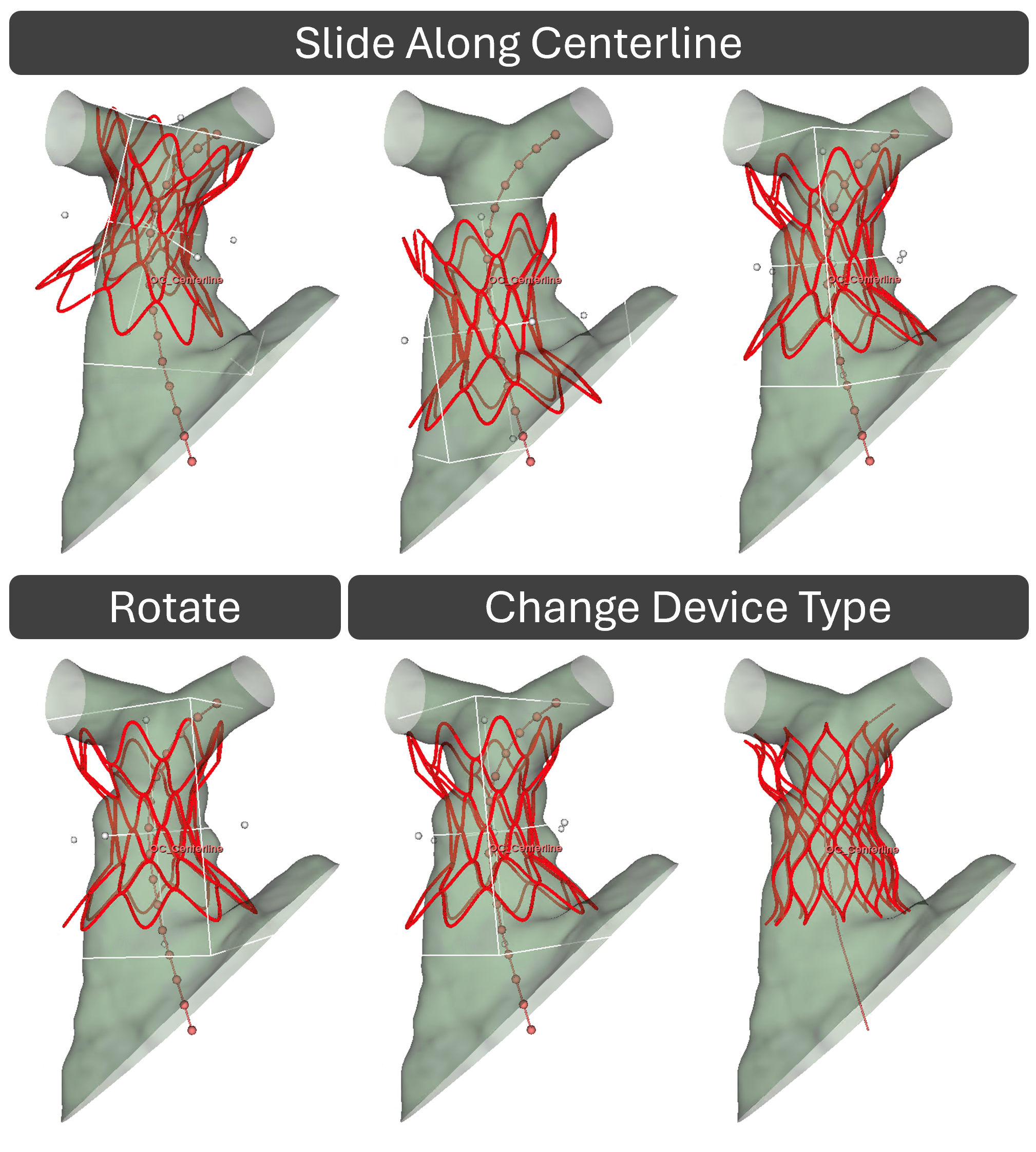}
    \caption{Device placement and TPVR simulation module functionality.}
    \label{fig:TPV_placement}
\end{figure}

Figure \ref{fig:TPV_placement} shows the procedure for device placement and the TPVR Simulation Module functionality. The TPVR Simulation module uses the vessel centerline and the central axis of the device to position the TPV in the RVOT. The vessel centerline was defined automatically using the Vascular Modeling Toolkit extension centerline extraction functionality or placed manually using an open curve in the Markups module \cite{piccinelli_framework_2009} in 3D Slicer. Once a device has been selected, the device is snapped to the centerline for precise placement. If the device imports upside-down with respect to the RVOT model and associated CT image, there is a button to flip the device into the correct orientation while keeping it aligned with the centerline. The user can slide the TPV along the centerline using a slider control or by inputting a value representing the desired location along the centerline. The centerline has been discretized into sections with values from 0 to 100, with 0 corresponding to the most distal point and 100 representing the proximal endpoint. As the device is slid along the vessel centerline, the angle of the device automatically adjusts so that its vertical axis is tangent to the vessel centerline at the center of mass of the TPV. In addition, the TPV can be rotated or translated freely to refine pre-simulation placement. This can be done via menu sliders controlling orientation and spatial position or by using the 3D interactive handler and dragging the device. Should the user wish to change their device type or centerline at any point in the process, they can do so by returning to the Device Selection or Device Positioning subsection, respectively, and selecting a different one. Doing so will not alter the location of the TPV unless they then choose to move it. The device positioning and selection process is highlighted in Video-1.

\subsubsection{Tube Creation for Simulation}
To simulate uniform compression of the device prior to TPV expansion, we created a curved tube surface that surrounds the device using the vessel centerline and the MarkupsToModel extension, leveraging vtkTubeFilter \cite{schroeder_visualization_2006}. The centerline curve points for the tube are interpolated using global least squares polynomial interpolation. The Centerline Parameters section of the module enables the user to toggle the visibility of the tube and set the radius and number of sides of the tube that capture the circumference of the tube. The tube’s mesh density depends on the number of points used to form the centerline and the number of sides. Increasing the number of sides adds more elements circumferentially. The module accommodates radii values of 0 to 50 mm. We chose a radius for the tube that was sufficient to contain the device at the desired position and angle without penetrating the tube, usually 28 mm. The number of elements spanning the tube vertically can be increased by resampling the centerline to have more points under Markups. Tube mesh density can be altered to prevent penetration of the device during simulations. Video-1 also highlights the tube creation process. The curved tube mesh was split into two halves for a staged release of the device inside the vessel as is further explained in Section 3.4.6.

\subsubsection{Export}
After the TPV was positioned and the tube given an appropriate radius and mesh density, we exported the tube, TPV, and RVOT shell from the Import / Export subsection of the TPVR Simulation module. The module applies a transform to all selected components upon export so that the saved files are aligned along the Z axis in global Cartesian coordinates. The Z axis corresponds to the vertical axis through the TPV. This facilitates the definition of local material axes and imposes constraints during simulation. The transformation can be undone by using the import feature when bringing the geometries back into 3D Slicer. The 3D Slicer scene was saved as a .mrb (medical reality bundle) file so it could be accessed later for visualization. 

\subsubsection{FEM Mesh Creation}
For each patient case, the shell mesh representing the inner surface of the RVOT was exported from 3D Slicer as a stereolithography (.STL) file and remeshed in Blender (V4.4) as quad elements. These surfaces were reduced in density and meshed as quads with an average element size of 1.5 mm. This quad-meshed RVOT surface was exported as a .ply file from Blender and subsequently imported into FEBio. The quad mesh was then extruded normally outward to have a thickness of 1.5 mm \cite{donahue_finite_2024, vanderveken_mechano-biological_2020} to create a hexahedral-dominant mesh for the entire RVOT due to better stability and accuracy in FE simulations over tetrahedral- or pentahedral-dominant meshes. The RVOT wall consisted of four solid mesh layers across its entire thickness to replicate the orthotropic nature of the artery based on the results of a mesh convergence study observing strain in increasing numbers of {wall mesh layers}. The final solid vessel geometries had between 24,000 and 32,000 elements, depending on the size of the vessel. The previously meshed TPV device .vtk and the tube .STL were brought into FEBio as well. The tube was converted to an editable mesh and assigned a shell thickness of 2.0 mm. 

\subsubsection{Simulation of Device Expansion}
\begin{figure}
    \centering
    \includegraphics[width=1.0\linewidth]{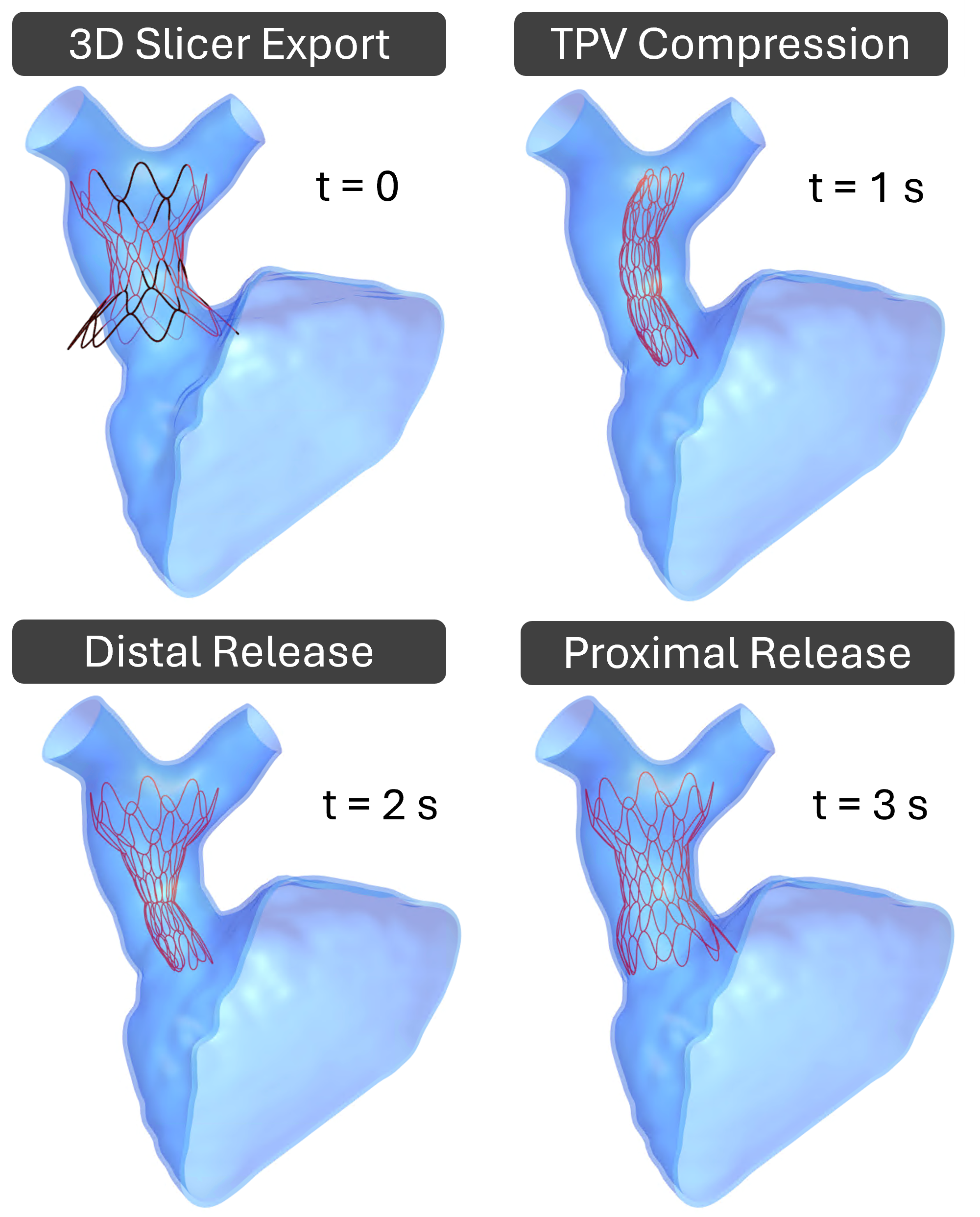}
    \caption{FEM modeling – staged deployment in FEBio.}
    \label{fig:Stent_staged_deployment}
\end{figure}

The simulation was divided into two analysis steps to capture the curved compression of the device, as though fitting it into the catheter delivery system, followed by the staged deployment in the vessel, as also shown in Figure \ref{fig:Stent_staged_deployment}. The distal half of the TPV device expands before the proximal half to mimic the delivery system sheath being pulled back and the device released. The tube was compressed over one second at a rate such that the enclosed device was well contained within the vessel. To achieve this, a displacement normal to the surface of the curved tube was applied to compress the device uniformly. Although the devices should fit within a delivery catheter system with a diameter of 8.33 mm (25 Fr) \cite{medtronic_tpv_2021}, we experienced instability in the models at very high compression due to complexities in the contact problem, so we used a larger target diameter of 16 mm after compression for this study. Since the device and its expansion is constrained, this larger target diameter would not have an effect the final deployed configuration as long as the fully compressed device completely fits inside the vessel without any intersections or penetrations. In most of our example cases, the compression rate was 20 mm/s, for the tube had an initial radius of 28 mm. When a greater initial radius was required for the tube to fully surround the device without penetration, the rate was adjusted accordingly to meet the same target radius. During TPV deployment, each half of the device was allowed to expand at the same rate at which it was compressed. The device compression and expansion simulation in FEBio is shown in Video-1.

The RVOT is incompressible and exhibits orthotropic material behavior at large deformations. We instituted an uncoupled Holzapfel-Gasser-Ogden (HGO) constitutive model for the RVOT to account for the fiber orientation and orthotropic nature of the RVOT vessels. The model uncouples the deviatoric and volumetric contributions in the strain-energy function:
\begin{equation}\Psi_{r} = \tilde{\Psi}_{r} \left( \tilde{C} \right) + U(J)\end{equation}
with
\begin{equation}\tilde{\Psi}_{r} = \frac{c}{2} \left( \tilde{ I_{1}} - 3 \right) + \frac{k_1}{2k_2}\sum_{\alpha=1}^{2} \left( \exp \left( k_2 \langle \tilde{E_{\alpha}} \rangle ^2 \right) - 1 \right)\end{equation} 
and the volumetric portion
\begin{equation}U(J) =  \frac{k}{2} \left( \frac{J^2 - 1}{2} - \ln J \right)\end{equation}

\(\tilde{I}_1 = {tr\tilde{C} }\), 
\(\tilde{I}_{4 \alpha} = a_{\alpha r} \cdot\tilde{C} \cdot a_{\alpha r}\), and 
\(\alpha = 1, 2\).
\(\tilde{C}\) is the right Cauchy-Green deformation tensor.

\begin{math}
E_{\alpha} = \kappa \left( \tilde{I}_1 - 3 \right) + ( 1 - 3 \kappa ) \left( \tilde{I}_{4\alpha} - 1 \right)
\end{math}
represents the fiber strain \cite{gasser_hyperelastic_2006, maas_uncoupled_2025}. For parameters, \(k_1\) is the fiber modulus, \(k_2\) is the fiber exponential coefficient, \(k\) is the bulk modulus, \(\kappa\) is the fiber dispersion, and \(c\) is the shear modulus of the ground matrix. \(J\) is the determinant of the deformation tensor or the local volume ratio. \(a_{\alpha r}\) (\(\alpha = 1, 2\)) represents the vectors along which the two fiber families are oriented. We used HGO material properties as defined in Donahue et al. \cite{donahue_finite_2022}. Local material axes were assigned to all vessel solid mesh elements. For each element, the local X-axis was positioned along the maximum curvature or circumference of the vessel, the local Y-axis was positioned along the minimum curvature or longitudinal direction, and the local Z-axis was positioned normal to the vessel surface.

We adopted the isotropic elastic material model \cite{maas_isotropic_2025} present in FEBio for the Nitinol TPV. The density, Young’s modulus \(E\), and Poisson’s ratio \(\nu\) must be defined. We chose known material properties for Nitinol \cite{vemury_behaviour_2019}. The Nitinol device had a density of 6.45e-06 kg/mm$^3$ and a Poisson’s ratio of 0.33. Nitinol exhibits superelasticity, which means that its stiffness transitions from a stiff austenite phase to a softer martensite phase to accommodate the large strains experienced during compression. When a stent expands after being unloaded, the material undergoes a reverse transformation from martensite back to austenite, resulting in an increase in stiffness as it returns to its original shape. As such, a step function load curve was applied to the Young’s modulus to simulate the transition of the effective stiffness in the device between the compression and release phases. The hyperelastic strain-energy function for the isotropic elastic material is given as:

\begin{equation}W = \frac{1}{2} \lambda ( tr \boldsymbol{E} )^{2} + \mu \boldsymbol{E} : \boldsymbol{E}\end{equation}
where \textit{\textbf{E}} is the Euler-Lagrange strain tensor. \(\lambda\) and \(\mu\) are the Lamé parameters. The material parameters and those in the strain-energy function share the following relationship:
\begin{equation}\mu = \frac{E}{2 (1 + \nu)}\end{equation}
\begin{equation}\lambda = \frac{ \nu E }{(1 + \nu) (1 - 2 \nu)}\end{equation}
where, \(E\) is Young's modulus and \(\nu\) is Poisson's ratio. The model reduces to a linear elastic material for small strains. 

We used an unconstrained Neo-Hookean material \cite{maas_neo-hookean_2025} for the tube that compresses the device. For simplicity, the tube was assigned parameter values identical to the TPV, but the Young’s modulus was kept constant at 40 GPa. The strain-energy function is
\begin{equation}W = \frac{\mu}{2} ( I_{1} - 3 ) - \mu \ln(J) + \frac{\lambda}{2} (\ln J)^2\end{equation}
where \(C\) is the right Cauchy-Green deformation tensor, \(I_1\) and \(I_2\) are the first and second invariants of \(C\), and \(J\) is the determinant of the deformation gradient tensor. The model captures the non-linear stress-strain response but reduces to classical linear elasticity under small strains and small rotations \cite{bonet_nonlinear_2008}. All parameter values used for each component of the simulation are listed in Table \ref{tab:material_parameters}. A summary of various parameter values selection for the individual components of the simulation and guidance for adapting are included in the supplementary material in Table A1.

\begin{table}[!ht]
    \centering
    \caption{Material parameters values for TPVR simulation components.}
    \label{tab:material_parameters}
    \begin{tabular}{ccc}
        \hline
        \textbf{Material Model} & \textbf{Material Parameter} & \textbf{Value} \\ \hline
        \multicolumn{3}{c}{\textbf{RVOT \cite{donahue_finite_2022, donahue_finite_2024}}} \\ \hline
        \multirow{7}{*}{Uncoupled HGO}
                  & $\rho$    & 1.02$\times 10^{-6}$ kg/mm$^3$ \\
                  & $c$       & 200 kPa \\
                  & $k_1$     & 13480 kPa \\
                  & $k_2$     & 1.06 \\
                  & $\gamma$  & 18.85 deg \\
                  & $\kappa$  & 0.33 \\
                  & $k$       & 1500 kPa \\ \hline
        \multicolumn{3}{c}{\textbf{Nitinol TPV Device (TPV-25, Alterra) \cite{vemury_behaviour_2019}}} \\ \hline
        \multirow{3}{*}{Neo-Hookean}
                  & $\rho$    & 6.45$\times 10^{-6}$ kg/mm$^3$ \\
                  & $E$       & 4.0$\times 10^{7}$ kPa \\
                  & $\nu$     & 0.33 \\ \hline
        \multicolumn{3}{c}{\textbf{Tube}} \\ \hline
        \multirow{3}{*}{Neo-Hookean}
                  & $\rho$    & 6.45$\times 10^{-6}$ kg/mm$^3$ \\
                  & $E$       & 4.0$\times 10^{7}$ kPa \\
                  & $\nu$     & 0.33 \\ \hline
    \end{tabular}
\end{table}

Boundary conditions and contacts were imposed on all geometries but varied by step. A potential-based contact formulation was used to constrain the TPV to stay within the tube during compression and staged expansion \cite{kamensky_contact_2018, maas_contact_2025}. The same contact formulation was used between the TPV and the vessel during deployment of the device within the RVOT. A zero Z-displacement boundary condition was imposed on the central nodes of the TPV to prevent sliding along the vertical Z axis. The ends of the vessel were fixed. Expansion of the TPV was controlled by a normal displacement boundary condition applied to the surrounding tube and by the enforced contact potential, mimicking the self-expanding behavior of the Nitinol devices. To improve the stability of the simulation, mass damping was assigned to the device as part of a false transient approach. Since the RVOT geometry already corresponded to the diastolic phase of the cardiac cycle, no pressure was applied on the luminal layer of the RVOT and hence, physiological pre-stress was omitted in the simulations. A full-Newton non-linear solver was used for the analysis. Additional details on the numerical methods, space-time discretization, and weak forms used in FEBio are included in \cite{maas_febio_2012}. The simulation files have been uploaded to the FEBio model repository (\url{repo.febio.org/permalink/project/136}).

\subsubsection{Visualization of Results}
Device and RVOT metrics were visualized by exporting the expanded device and vessel from FEBio and importing them in 3D Slicer using the import section of the TPVR simulation module. After importing the mesh into Slicer, metrics calculated by FEBio were pre-processed so that they could be visualized on the vessel and model surface mesh. For vector metrics such as displacement, we computed the length of the displacement vector and visualized the magnitude of the vector on each point in the mesh. In metrics such as stress, we computed and visualized the absolute magnitude of the eigenvalues for each polygon in the mesh. The metric pre-processing generated single-component values for the metrics that were visualized by applying a scalar overlay on the mesh, coloring the surface based on the metric value.

To visualize the contact between the device and vessel, we calculated a B-spline transformation using the ScatteredTransform extension in 3D Slicer. The B-spline transformation reflects the deformation of the device mesh from its original location to its final position at the end of the FEBio simulation. We then applied the same deformation to a mesh of the fabric for the corresponding device. The distance between this deformed fabric mesh and the vessel wall was calculated using the ModelToModel distance extension in 3D Slicer. The absolute distance from the points in the device fabric mesh to the closest point on the vessel wall was calculated for each point in the fabric mesh, and vice versa. To visualize regions of contact, we applied a color gradient to the model representing the distance between the fabric and vessel mesh. We also hid or reduced the opacity of regions that were not within a specified distance threshold (1.5 mm) to highlight the regions of contact between the device fabric and the vessel wall.

\subsubsection{Quantitative Metrics of Compression}
Quantitative compression metrics included the 1st principal Lagrangian strain, 1st principal stress, and percentage contact area of the valve frame in the proximal and distal halves of the vessel at complete device expansion. The 99th percentile, 75th percentile, and maximum 1st principal Lagrangian strain and stress were plotted for each device and patient. The percentage of the total device area in contact with the vessel wall was calculated for the proximal and distal halves to determine the proximal and distal seal for each device.

\section{Results}\label{sec3}

The time taken for automatic segmentation of the RVOT models was 5 seconds. Manual corrections of the segmentations took approximately 5 minutes. Time taken for device placement was 45 seconds. Time taken to set up simulations in FEBio was 5 minutes. The average computational time was 2 hr 17 min for simulations with the TPV-25 and 5 hr 34 min for the Alterra on an Intel Core i9 10980 XE CPU @ 3.00GHz processor with 64.0 GB RAM system. The Alterra model required more elements than the TPV-25 due to a more complex shape, which contributed to an elevated runtime. All times are reported as averages across all three RVOTs and TPV devices.

The distribution of stress, strain, and stent-vessel contact is critical for evaluating device fit, mechanical stability, and potential risk of complications. High local stresses and strains in the vessel wall can indicate areas at risk for tissue injury, vessel remodeling, or even rupture, particularly important in fragile or surgically repaired RVOTs. Conversely, insufficient mechanical interaction (i.e., low strain or minimal contact) may suggest suboptimal anchoring of the device, increasing the risk of stent migration or paravalvular leak. To evaluate device fit, we examined the 1st principal strain and stress, as well as areas of contact between each stent and vessel. All quantitative metrics of compression have been depicted in Video-2. We divided the stent centrally about its vertical axis into proximal and distal halves in order to assess proximal and distal device compression separately.

Perimeter-derived diameters, as calculated in \cite{mcelhinney_transcatheter_2024}, for RVOT1, RVOT2, and RVOT3 when the TPV25 device was used were 30.72 mm, 28.58 mm, and 27.66 mm, respectively. Similarly, perimeter-derived diameters for RVOT1, RVOT2, and RVOT3 when the Alterra device was used were 31.44 mm, 28.91 mm, and 29.46 mm, respectively.

\begin{figure}
    \centering
    \includegraphics[width=1.0\linewidth]{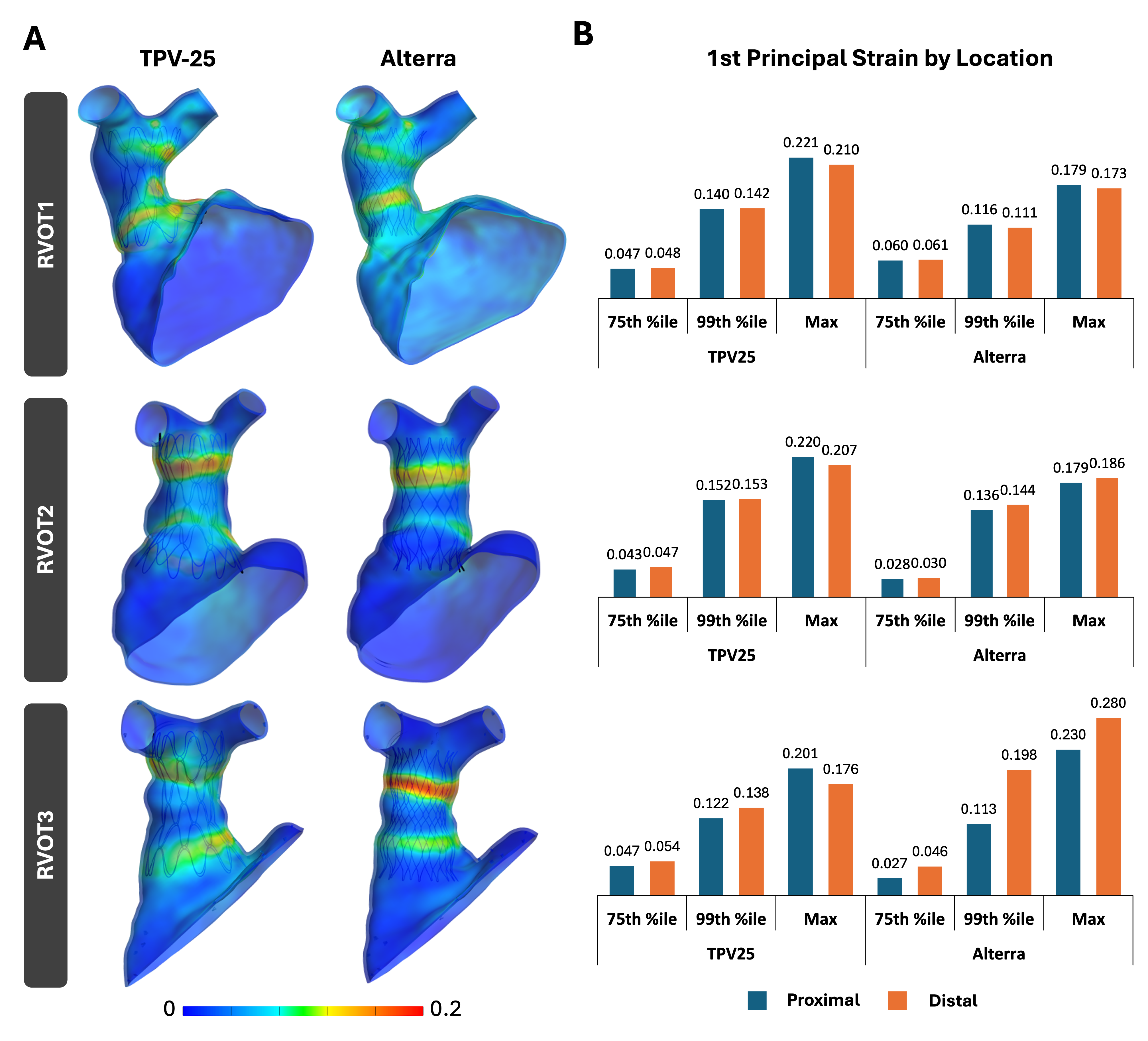}
    \caption{A) 1st principal Lagrangian strain distribution in simulated RVOT and TPV cases, and B) peak strain values observed in the RVOT across simulated cases, separated by proximal and distal halves.}
    \label{fig:results_strain}
\end{figure}

\begin{figure}
    \centering
    \includegraphics[width=1.0\linewidth]{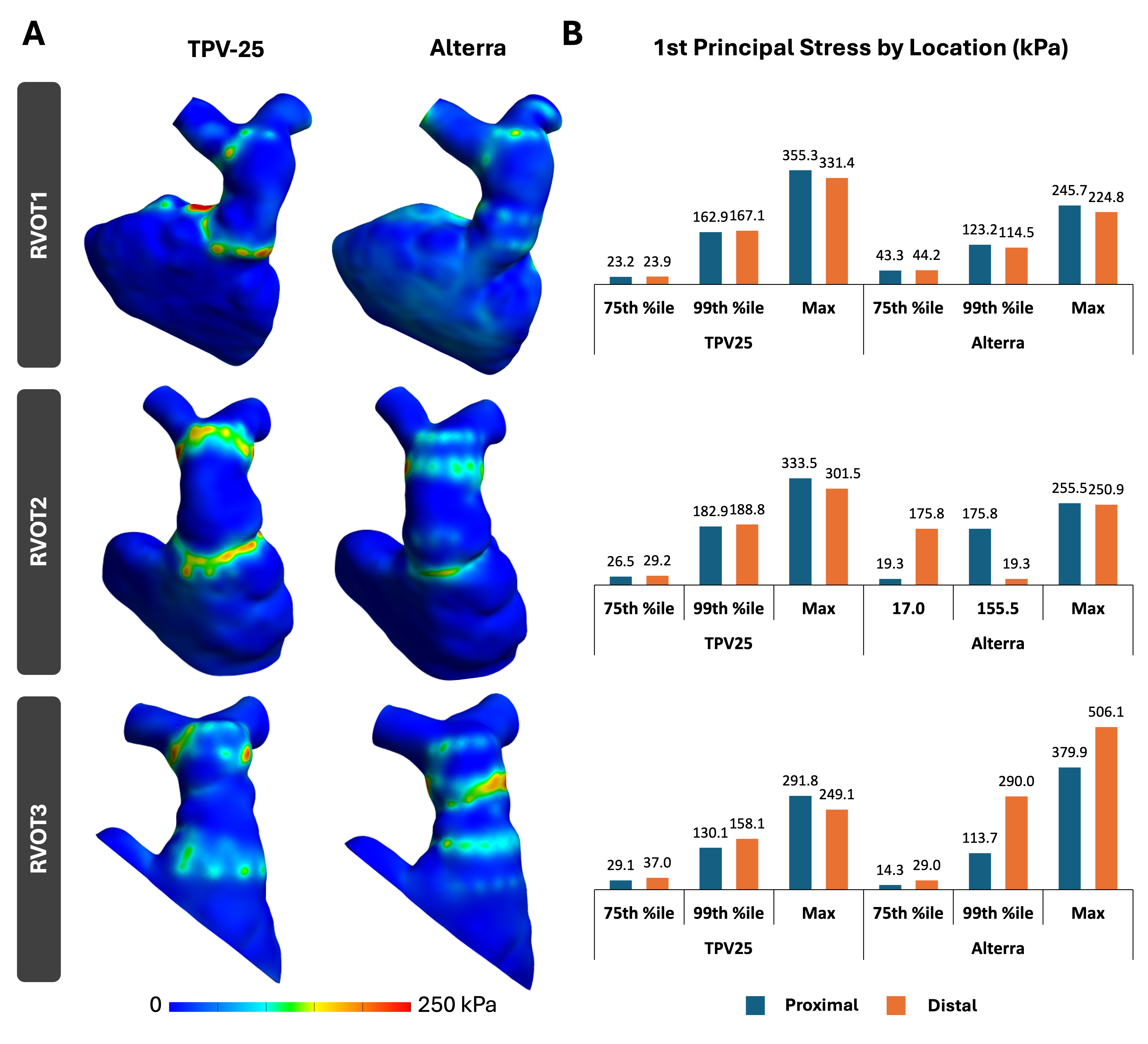}
    \caption{A) 1st principal stress distribution in simulated RVOT and TPV cases, and B) peak stress values observed in the RVOT across simulated cases, separated by proximal and distal halves.}
    \label{fig:results_stress}
\end{figure}

Figures \ref{fig:results_strain} and \ref{fig:results_stress} highlight the 1st principal Lagrangian strain and 1st principal stress distribution across the three RVOT models when the TPV-25 and Alterra devices are fully expanded, as well as the 99th percentile, 75th percentile, and maximum values of those strains and stresses. Areas of high stress and strain were observed at the proximal and distal ends of the RVOTs, where the device wireframes contact the RVOT vessel wall and where the device geometries are the broadest. 

Across these three RVOT models, the 75th percentile strain and stress values were fairly similar between the two stents. For RVOT1, the TPV-25 resulted in greater 99th percentile and maximum strain and stress than the Alterra. For both stents, strain and stress at the proximal and distal halves of the RVOT were similar. For RVOT2, like RVOT1, the TPV-25 produced a greater 99th percentile and maximum strain and stress than the Alterra. Proximal and distal strains and stresses were similar. In RVOT3, trends varied from RVOTs 1 and 2. At the proximal half, TPV-25 resulted in greater 99th percentile strain and stress, but lesser maximum strain and stress than the Alterra. At the distal half of the RVOT, TPV-25 produced a lower 99th percentile and maximum strain and stress than the Alterra. On comparing the two halves, the distal half of RVOT3 showed greater strain and stress values than the proximal half across both stents, except for the maximum stress when the TPV-25 was used. 

\begin{figure}
    \centering
    \includegraphics[width=1.0\linewidth]{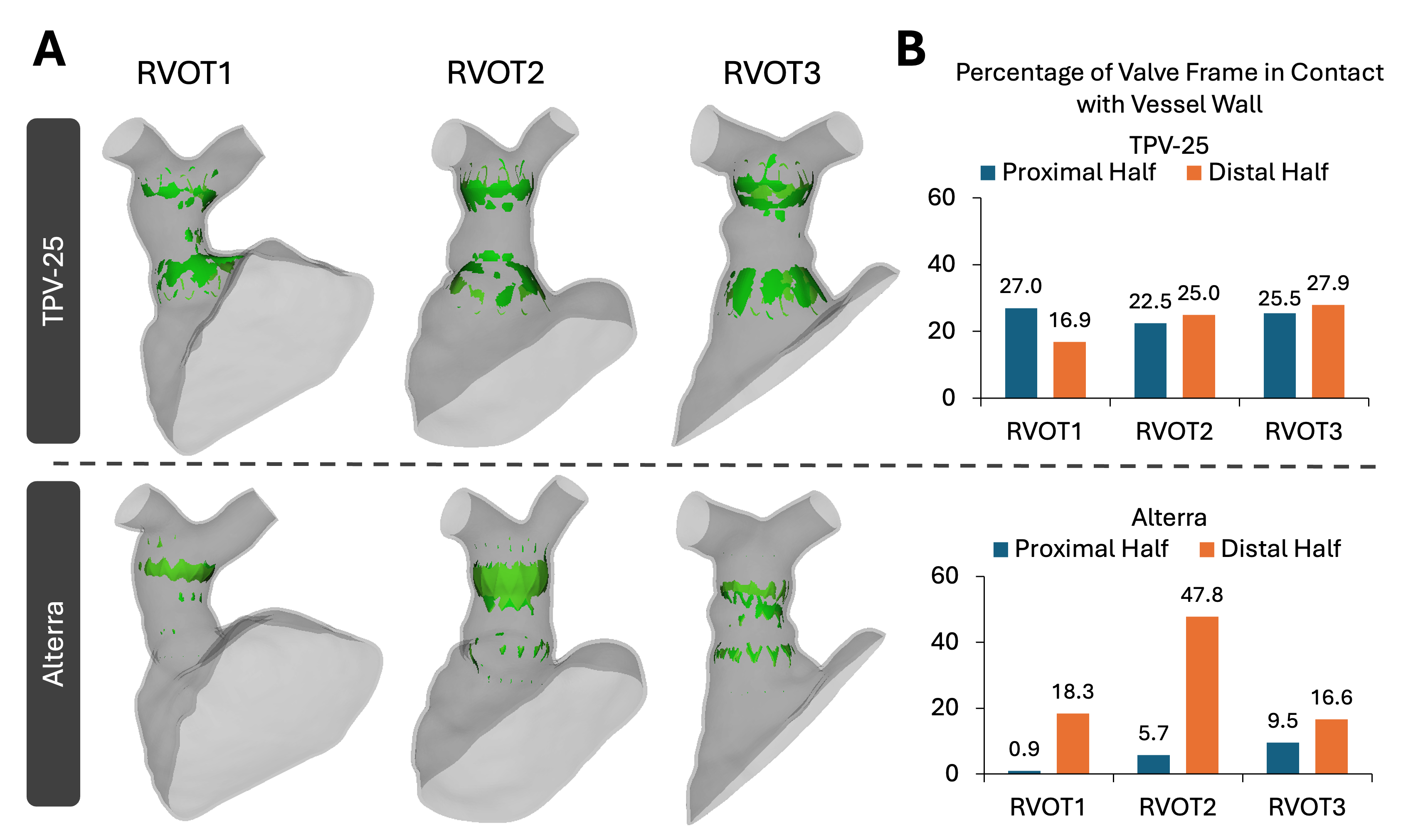}
    \caption{Contact visualization for simulated RVOT and TPV cases, A) Visualization of the contact area on RVOT wall across all simulated cases, B) Bar plot depicting the percentage of the stent/valve frame in contact with the vessel wall, separated by proximal and distal halves.}
    \label{fig:results_contact}
\end{figure}

Contact maps generated for each RVOT and stent condition are shown in figure \ref{fig:results_contact}. Greater contact was observed at the proximal and distal ends of the RVOT where the stent geometries are the broadest. Overall, the TPV-25 had a greater percentage of the total stent area in contact with the vessel walls than the Alterra. On analyzing by location, the TPV-25 had fairly similar contact percentages in the proximal and distal halves. However, for RVOTs 1 and 2, the Alterra device contacted only ~1\% and ~6\% of its proximal area of the stent in the proximal half. For the Alterra stent, contact percentages in the distal half were much greater than those in the proximal half, unlike the TPV-25 stent.

By quantifying and comparing stress, strain, and contact patterns across multiple patient-specific anatomies and devices, this study underscores and reaffirms the importance of personalized anatomical reconstruction in TPVR \cite{swanson_3d_2024}. The observed variability in these metrics across different RVOTs, between stents (TPV-25 vs. Alterra), and locations in the RVOT (proximal vs. distal) highlights that no single device performs optimally in all anatomies, reinforcing the need for simulation-driven preprocedural assessment.

\section{Discussion}\label{sec4}

In this study, we developed an integrated open-source workflow to simulate TPV deployment in patient-specific, image-derived RVOT anatomies, with the long-term goal of improving patient selection and device fit for TPVR. The workflow incorporates automated RVOT segmentation, interactive TPV selection and placement through a dedicated SlicerHeart module, followed by simulation of device–tissue interaction using FEBio. We evaluated two commonly used TPVs across three distinct patient anatomies, assessing first principal stress, strain distributions, and regions of contact to determine device fit to establish an initial paradigm that can be further developed by the research community. Although the individual building blocks of our workflow draw on established methods, the novelty of our open-source workflow is the integration itself: a single, open-source pipeline that carries a patient from a CT scan through automatic segmentation, interactive device selection and positioning, staged finite element deployment of a self-expanding device, and quantification and visualization of the results on the original anatomy. Additionally, this open-source workflow allows investigating anatomies of patients for whom commercial modeling and screening tools are not necessarily developed due to their small numbers, and hence, low commercial market incentives. To the authors' knowledge, no prior studies connect these stages end-to-end in publicly available, open-source tools. 

While image-based simulations are increasingly used to support adult structural interventions, similar workflows have yet to be routinely adopted in congenital heart disease. \cite{xuan_stent_2017, brown_patient-specific_2023, grossi_validation_2024} However, patient-specific modeling may be especially valuable in this setting due to the high anatomical variability and small patient cohorts, which limit the feasibility of large clinical trials and empirical design validation.

We begin to address that gap by simulating the deployment of commercially available self-expanding TPVs in a range of pediatric RVOT anatomies. Using the HGO model to represent the RVOT wall, we demonstrate the feasibility of preprocedural prediction of device expansion, positioning, and vessel interaction in patient-specific cases. To further enhance the practicality of our workflow, we incorporated machine-learning-based segmentation for automated RVOT extraction. This capability greatly improves efficiency and supports the trajectory toward future clinical applicability of patient-specific modeling \cite{barak-corren_image-derived_2025}.

Comparing the stress, strain, and contact patterns across the three RVOTs and two stents explored in this study highlights that no single device performs optimally in all anatomies. Although the Alterra device results in a smaller percentage contact area of the stent compared to the TPV-25, this can be attributed to the design of the device. The Alterra device has tines at its proximal and distal ends, which embed into the RVOT wall on installation. Scar tissue developing around the area where the stent's tines impinge the RVOT wall seemingly results in strong placement of the device. However, this behavior is not captured in our simulations. Associating the stress, strain, and contact patterns for a cohort of patients for whom this procedure was successful or unsuccessful would provide data to link these mechanical measures to actual clinical outcomes. 

Only a relative comparison of metrics across patient RVOTs and devices was conducted. Direct, physiologically meaningful validation is not currently feasible in the population of subjects explored in this study. Post-implantation CT scans are not part of routine care for these subjects and is generally obtained only when a patient develops a complication. The available ground-truth cases are, therefore, rare and systemically biased toward adverse outcomes. Routine post-implant CT in asymptomatic patients also exposes them to unjustifiable radiation exposure. For the Alterra device, post-implantation CT is more often available, but a balloon-expandable SAPIEN valve is subsequently implanted within the Alterra landing zone, which alters the deployed geometry relative to the self-expanding deployment in a native outflow tract that our framework models. Post-implant imaging in that setting, therefore, does not provide a valid comparison our simulations. This open-source workflow is not meant to not solve every challenge required to be used clinically. Validation and regulatory alignment must be established before such a framework can inform clinical decisions. 

Despite the strengths of our integrated approach, including automatic segmentation, TPV selection, and direct export to FEBio for soft tissue interaction modeling, several limitations remain. First, the simulations focus solely on biomechanical outcomes and do not incorporate blood flow or hemodynamic forces. Future work will incorporate fluid–structure interaction (FSI) modeling to enable more comprehensive predictions \cite{black_open-source_2026}.

Device–tissue interaction was modeled using a contact potential method, based on the formulation by Kamensky et al. \cite{kamensky_contact_2018}, which applies a repulsive force when surfaces come within a specified distance. Although this approach maintains an artificial small gap between the stent and vessel wall, it prevents penetration and offers a more stable solution compared to the sliding-elastic formulation in FEBio, especially for soft tissue interactions \cite{maas_febio_2012}. Future exploration of the optimal contact for this application is needed.

Notably, the material phase transformation in the Nitinol stent is influenced by the progression of the martensitic volume fraction, which changes in response to internal stress during deformation. Although our step-function approach provides a reasonable approximation of the overall characteristics of the stent, it does not fully capture the mechanical hysteresis behavior associated with the transformation. Subsequent work will aim to implement a more physically representative superelasticity model \cite{auricchio_shape-memory_1997} based on the martensite fraction, similar to those presented in previous literature \cite{auricchio_carotid_2011, tzamtzis_numerical_2013, maleckis_nitinol_2018}. A more representative material model for the stent may result in different softening and stiffening mechanisms that could influence the final expansion of the stent, and the strain and stress values and distribution in the RVOT. However, this needs to be explored further.

Simulation run times were relatively long due to the use of solid elements in modeling the TPV device. As a next step, we plan to integrate 1-dimensional beam elements, which are expected to reduce computational cost while more effectively capturing the bending behavior of the stent.

In our simulations, the RVOT wall was modeled using HGO parameters from existing literature \cite{donahue_finite_2022}, without accounting for layer-specific material properties due to the lack of experimental data in pediatric tissues. Currently, tissue-derived material data for the pulmonary artery are limited to a small number of patients, either infants \cite{cabrera_mechanical_2013} or older adults \cite{holzapfel_determination_2005}. The mechanical behavior of tissues based on material parameter values from \cite{cabrera_mechanical_2013} will not necessarily be representative of the behavior of tissues of the adolescent test subjects in this study. However, they are still the most relevant material properties available in literature. Non-conduit RVOTs, which include both native (non-patched) and surgically patched RVOTs, have anatomies that are categorized by substantial heterogeneity in geometry and biomechanical properties. These include unpredictable tissue compliance and distensibility. A comprehensive sensitivity analysis of RVOT material parameters may help understand the variation observed in RVOT behavior due to device expansion. Given the difficulty of in vivo mechanical characterization and the expected interpatient variability of vessel wall parameters, a sensitivity analysis could be performed to better quantify and interpret the variability in the mechanical response of the RVOT anatomy during interaction with TPVR devices. Age-related modeling, surgical history (e.g. patch implantation), and long-term pathological adaptation are likely to significantly alter tissue stiffness and anisotropy, further corroborating the need for a preliminary sensitivity analysis. Simulation accuracy will continue to improve as new tissue mechanical testing data and novel measurement techniques become available \cite{wu_noninvasive_2025}.

CTA images, and hence, the RVOT geometries of all three patients, corresponded to the diastolic phase. Prior literature has reported that the phase of the CT dataset is crucial for assessing device stability \cite{sturla_planning_2025}. While the use of the diastolic phase for analysis may have a negligible impact in the context of conduit RVOTs, its influence in non-conduit RVOTs, where compliance and dynamic dimensional changes are more pronounced, could be more substantial and needs to be explored further. Since the RVOT and pulmonary artery are dynamic, future simulations will include both diastolic and systolic models to allow comparison over the cardiac cycle. While the RVOT geometry reconstructed from CT imaging acquired during the diastolic phase, the omission of physiological pre-stress conditions may affect the mechanical response of the assumed RVOT vessel wall, potentially influencing its stress-strain behavior.

Another limitation of this study is the small sample size and lack of validation against clinical post-implantation geometries. Typically, post-implantation CTA scans in patients undergoing these procedures are performed only if there is a device functionality failure, which limited our ability to perform this direct comparison. Future work will include a larger cohort and comparisons with post-procedural imaging to evaluate predictive accuracy.

\section{Conclusion}\label{sec5}
The open-source and adaptable nature of our platform makes it well-suited for further exploration and development by the research community. While it provides a feasibility assessment for TPVR, further development and rigorous validation are needed before potential clinical application. Nevertheless, by enabling the development and testing of patient-specific workflows, our platform provides a foundation for evaluating both novel devices and new uses of existing technologies \cite{tsao_transcatheter_2025}. With ongoing refinement and validation, this framework has the potential to become a valuable clinical tool for optimizing device selection and guiding procedural planning in TPVR and beyond.

\backmatter

\bmhead{Acknowledgements}
This work was supported by NIH R01HL153166, K25HL168235, 2R01GM083925, a CHOP Cardiac Center Innovation Award, the Topolewski Pediatric Valve Center at CHOP, and a CHOP Research Institute Post-Frontier Award.

\section*{Declarations}
Matthew J. Gillespie is a consultant for Medtronic. The other authors have no disclosures. 

\bibliography{references.bib}

\end{document}


\begin{figure}
    \centering
    \includegraphics[width=1.0\linewidth]{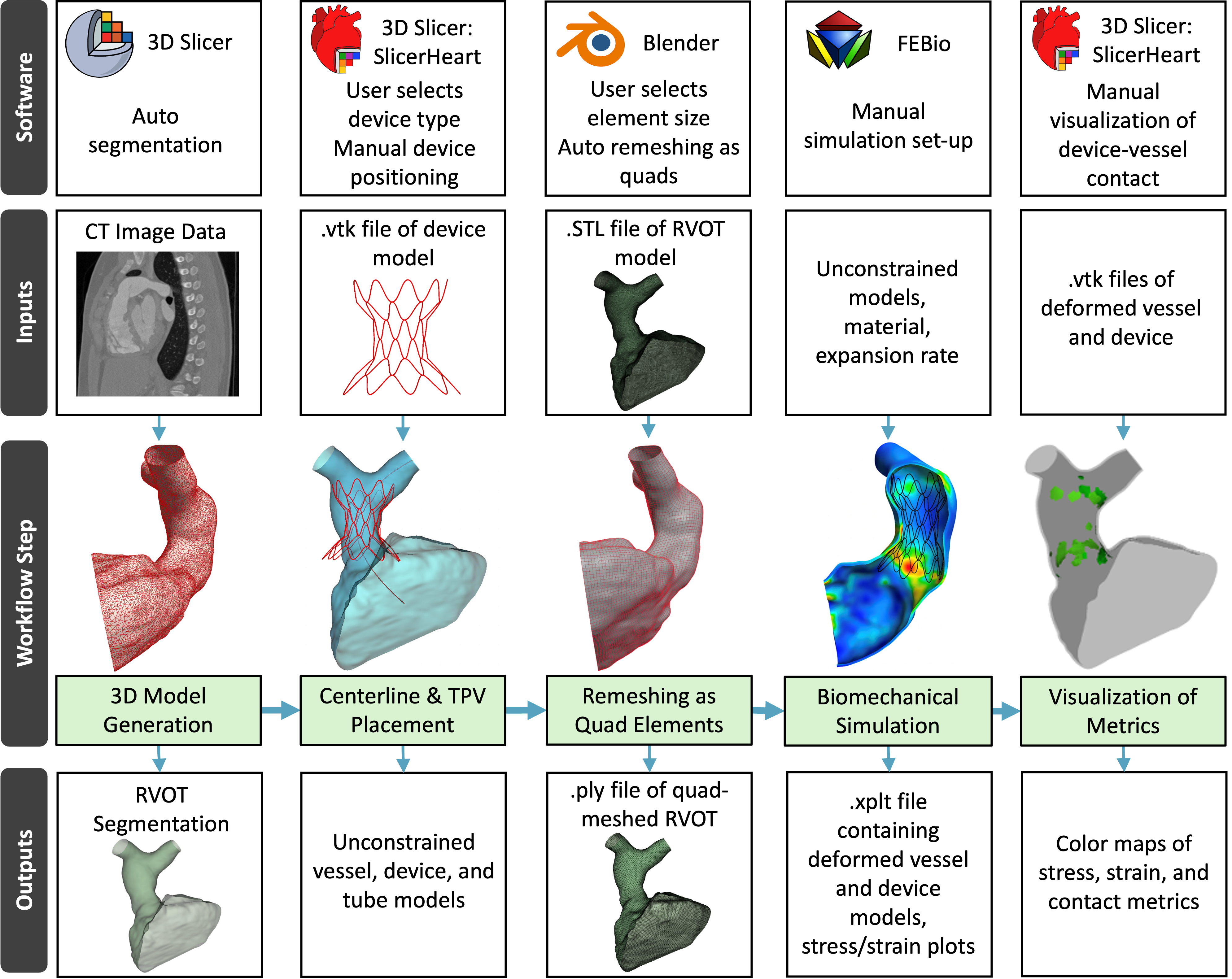}
    \caption{Detailed overview of modeling pipeline}
    \label{fig:workflow_detailed}
\end{figure}

\begin{table}[!ht]
    \centering
    \caption{Summary of key simulation parameters, sources, and guidance for adaptation.}
    \label{tab:simulation_parameter_summary}
    \begin{tabular}{m{3.2cm} m{2.8cm} m{3cm} m{3.5cm}}
        \hline
        \textbf{Parameter} & \textbf{Value used} & \textbf{Source / Rationale} & \textbf{Guidance for adapting} \\ \hline
        \multirow{7}{=}{RVOT uncoupled HGO material parameters}
            &  Density $= 1.02\times10^{-6}$ kg/mm$^3$ & \multirow{7}{=}{Donahue et al. 2022, 2024}
            & \multirow{7}{=}{Substitute case-specific values if available; companion sensitivity analysis reports each parameter's influence} \\
            & $c = 200$ kPa & & \\
            & $k_1 = 13480$ kPa & & \\
            & $k_2 = 1.06$ & & \\
            & $\gamma = 18.85$ deg & & \\
            & $\kappa = 1500$ kPa & & \\ \hline
        \multirow{4}{=}{Nitinol device Neo-Hookean material parameters}
            & Density $= 6.45\times10^{-6}$ kg/mm$^3$ & \multirow{4}{=}{Vemury et al. 2019}
            & \multirow{2}{=}{Device-dependent; change to match device material} \\
            & $E = 40$ GPa & & \\
            & $\nu = 0.33$ & & \\ \hline
        \multirow{4}{=}{Compression tube Neo-Hookean material parameters}
            & Density $= 6.45\times10^{-6}$ kg/mm$^3$ & \multirow{2}{=}{Numerical compressor; no physical meaning}
            & \multirow{2}{=}{Keep stiff relative to the device} \\
            & $E = 40$ GPa & & \\
            & $\nu = 0.33$ & & \\ \hline
        Target compressed stent diameter & 16 mm & Contact stability & Any value that fully contains the compressed device without penetration \\ \hline
        Compression tube initial radius & 28 mm (module allows 0 to 50 mm) & Contains device without penetration & Adjust to device and vessel size \\ \hline
        Tube compression/expansion rate & 20 mm/s & Quasi-static (false transient approach) & Rate-independent at convergence; reduce if unstable \\ \hline
        Mass damping coefficient & 2000 & False-transient stabilization & Increase if solution is oscillatory; does not affect final equilibrium \\ \hline
        RVOT wall: thickness, layers, element size & 1.5 mm thickness; 4 mesh layers; 1.5 mm elements & Mesh-convergence study (Section 3.4.5) & Re-run convergence check for a new RVOT geometry \\ \hline
        Metrics reporting percentiles/contact map threshold & 99th, 75th, and maximum; threshold = 1.5 mm & Reporting choice; contact threshold based on potential contact radius & Adjust thresholds to suit the intended analysis \\ \hline
    \end{tabular}
\end{table}